# Single Cycle Thin Film Compressor
## Opening the door to Zeptosecond-Exawatt Physics

G. Mourou[1, 3], S. Mironov[2], E, Khazanov[2, 3], A. Sergeev[2]

1) IZEST, Ecole Polytechnique, 91128 Palaiseau, France
2) Institute of Applied Physics RAS, 46 Ul'yanov Street, 603950, Nizhny Novgorod, Russia
3) University of Nizhni Novgorod, 23 Prospekt Gagarina, 603950, Nizhny Novgorod, Russia

## Abstract

*This article demonstrates a new compression scheme that has the potential to compress a high energy pulse as high as a few hundred Joules in a pulse as short as one optical cycle at 0.8μm making a true ultra-relativistic $\lambda^3$ pulse. This pulse could have a focused intensity of $10^{24}$W/cm$^2$ or $a^0$ of 1000. It could form an efficient, 10%, relativistic mirror that could compress the pulse to the atto-zeptosecond regime, with an upshifted wavelength of 1-10keV. This technique could be a watershed making the entry of petawatt pulses into the exawatt and zeptosecond regime possible.*

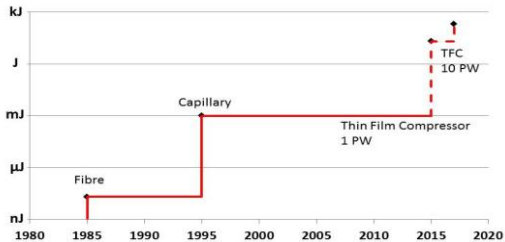

Fig. 1 Evolution of few optical cycle pulses over the years

There is a tendency to think that ultrashort pulse is the apanage of small scale laser. In the pulse duration-peak power conjecture [1] the opposite was demonstrated. Pulse duration and peak power are entangled. To shorten a pulse, it is necessary first to increase its peak power. In this article we show an example that illustrates this prediction, making possible the entry of laser into the zeptosecond and exawatt domain.

Since the beginning of the 1980's optical pulse compression [2] has become one of the standard ways to produce femtosecond pulse in the few cycle regime. The technique relies on a single mode fiber and is based on the interplay between the spectrum broadening produced by self phase modulation and the Group Velocity Dispersion necessary to stretch the pulse. The combination of both effects contributes to create a linearly frequency- chirped pulse that can be compressed using dispersive elements like grating pairs, prism pairs or chirped mirrors. In his pioneering experiment Grisckkowsky et al.[2] used a single mode optical fiber and were able to compress a picosecond pulse with nJ energy to the femtosecond level. This work triggered an enormous interest that culminated with the generation of a pulse as short as 6fs corresponding to 3 optical cycles at 620nm by C.V Shank's group [3] see Fig. 1. In their first experiment the pulse was only 20nJ, clamped at this level by the optical damage due to the core small size.To go higher in energy O. Svelto and his group [4] introduced a compression technique based on fused silica hollow-core capillary, filled with noble gases and showed that they could efficiently compress their pulses to the 100μJ level. Refining this technique O. Svelto, F. Krausz et al [5] could compress a 20fs into 5 fs or 2 cycles of light at 800nm the energy was typically sub mJ, see Fig. 1. In both cases, like with single mode fiber, the compression effect was still driven by the interplay between self phase modulation and group velocity dispersion.

To go higher in energy, bulk compression was attempted by Corkum and Rolland [6]. In their embodiment the pulse is free-propagating and not guided anymore. The pulse was relatively long around 50fs with an input energy of 500μJ leading to an output pulse of 100μJ in 20fs. This scheme is impaired by the beam bell shape intensity distribution. It leads to a non-uniform broadening compounded with small scale self focusing making the pulse impossible to compress except for the top part of the beam that can be considered as constant limiting the efficiency and attractiveness of this technique.

**Large energy pulse compression: Thin Film Compression (TFC), Fig.2**

Here we are describing a novel scheme capable to compress 25fs large energy pulses as high as 1kJ to the 1-2 fs level. We call this technique Thin Film Compressor or TFC. As shown in our simulation this technique is very efficient >50% and preserves the beam quality.

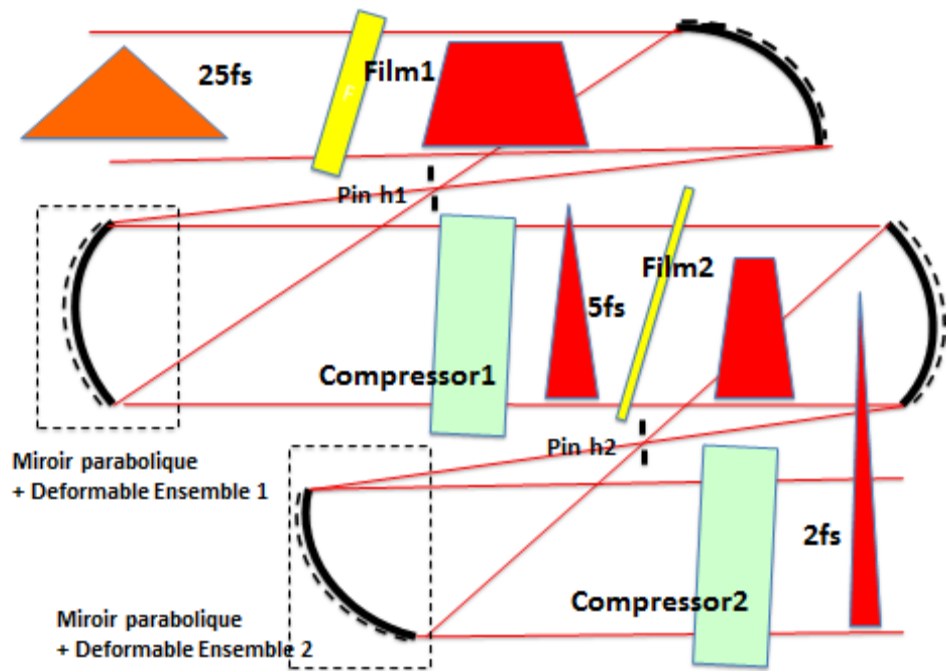

Fig.2 Embodiment of a double Thin Film Compressor TFC Thin film"plastic" of 500μm thickness as uniform as possible is set in the near field of PW producing a flat-top beam with a B of about 3-7. The beam propagate through a telescope composed of 2 parabolae, used to adjust finely the B and reduce the laser beam hot spots. Before compression the beam is corrected for the residual wavefront non-uniformity of the beam and the thin film thickness variations. The pulse is compressed using chirped mirrors to 6.4fs. The measurement performed using a single shot auto-correlator. The same steps is repeated in a second compressor with a film of 100 μm producing an output of 2 fs, 20J.

Unlike in the previous bulk compression technique performed with large scale laser exhibiting bell shape distribution, the technique relies on the top hat nature of large scale femtosecond lasers when they are well constructed. The Fig. 3a shows the output of a PW laser generating 27J in 27fs called CETAL in the Institut National de Laser, Plasma et Radiophysique (INFLPR) in Bucarest. Similar flat top energy distributions are exhibited by the Bella system at Lawrence Berkeley Laboratory. The next generation of high power laser will deliver 10PW like ELI-NP in Romania or Apollon in France, with a similar top hat beam. Our simulation shows that the pulse being already very short, i.e 27fs will require a very thin optical element of a fraction of a mm thick for a beam of 16cm diameter. This element will be extremely difficult to manufacture, extremely fragile to manipulate and very expensive, making the idea of pulse compression of high energy pulse unpractical. Our solution is to use a thin "plastic" film of ~500μm with a diameter of 20cm. The element, that we call plastic for simplicity could be polymer thermoplastic amorphes, like the PVdC, the additived PVC, the triacetate of cellulose, the polyester, or other elements as long as they are transparent to the wavelength under study, robust, flexible and exhibit a uniform thickness, ideally within a fraction of a wavelength. It is paramount to have a thickness as uniform as possible across the beam but does not have to be flat. As opposed to a thin (fraction of a mm) quartz, silicate over a dimension of 20cm, it is abundant, inexpensive and sturdier. It should be susceptible to withstand the laser shot without breaking. In case where the film breaks, it can be replaced, cheaply, easily for the following shot. In the preferred embodiment shown in fig. 2, the laser beam is focused by an off axis parabola with a f# about 10. The focused beam plays two roles. a) it can be used to adjust the beam intensity by sliding the film up and down (over a small travel though) in order to optimize intensity and b) to provide a means to eliminate the high spatial frequencies produced by the beam nonuniformities due to the small scale focusing. A pinhole of suitable dimension is located at the focus. After the focal point the beam is re-imaged to infinity by a second parabola. The pulse can be measured at this point using a standard single shot auto-correlator technique. Simulations, in the next chapter demonstrate the possibility to compress a 27J, 27 fs into 6fs in a first stage and 2fs in a second stage where the plastic thickness is 100μm. The beam remains of good quality after this double compression as shown in fig.3.

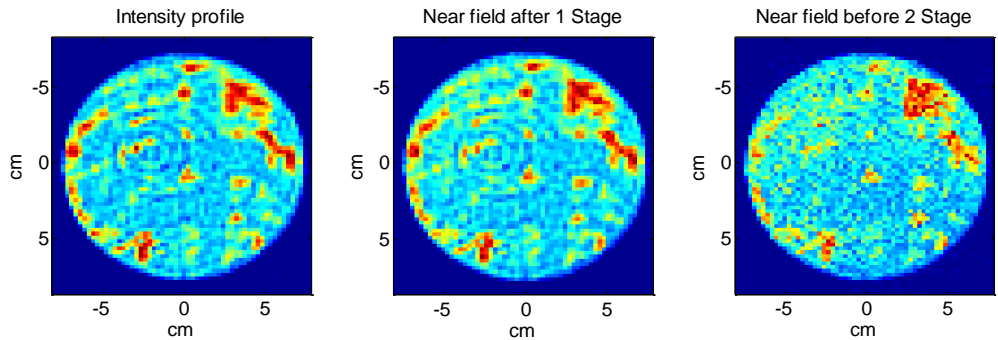

Fig. 3 This figure shows the intensity across the beam profile: a) at the laser output, b) after the first stage (no spatial filter, c) after the second stage (no spatial fiter)

Because there is no real loss in the system we expect an overall compressor efficiency close to >50%. As a consequence the peak power is increased close to 10 times. Note that ideally, after each "thin films" a wave front corrector is installed to take into account a possible non-uniformity of the film thickness that could not affect the B but would be harmful to the wave front. This simple technique provides a spectacular reduction in pulse duration of more than 10 time transforming a PW laser into a greater that 10PW laser. It can also be extended to the 10PW regime to boost its power to more than 100PW or .1EW.

**Modelling of the two stage Thin-Film-Compressor**

Let's consider the physical model thoroughly. The main process which is responsible for spectrum broadening in solid materials is self-phase modulation. The self-phase modulation is a result of changing refractive index at intense radiation:

$$n = n_0 + 1/2 \cdot n_2 \cdot |A|^2 = n_0 + \gamma \cdot I$$

Here $A(t-z/u, z)$ − complex amplitude of electric field, $I$ – intensity, $n_0$ – linear part of refractive index, $\gamma [cm^2/kW] = (2 \cdot \pi / n_0)^2 \cdot \chi^3 [ESU]$, $\chi^{(3)}$ − nonlinear susceptibility. Typical values of $\gamma$ for optical glasses are $(3 \div 8) \cdot 10^{-7} cm^2/GW$ [7]. The other important phenomenons are linear disperion – the dependence of refractive index versus wave length and effect of self-steepening. The influence of the processes on pulse parameters can be described in the frame of quasi-optical approximation [8]:

$$\frac{\partial A}{\partial z} + \frac{1}{u}\frac{\partial A}{\partial t} - i\frac{k_2}{2}\frac{\partial^2 A}{\partial t^2} + i\gamma_1 |A|^2 A + \frac{3\pi \cdot \chi^{(3)}}{n_o \cdot c}\frac{\partial}{\partial t}\left(|A|^2 A\right) = 0$$

Here, $\gamma_1 = (3\pi \cdot k_0 \cdot \chi^{(3)})/(2 \cdot n_o^2)$, $u$ – group velocity, c – speed of light, t – time, z – longitudinal coordinate, $k_2 = \left.\dfrac{\partial^2 k}{\partial \omega^2}\right|_{\omega_o}$ – parameter of group velocity dispersion (GVD) and $k_0$ – wave vector.

We use the equation with the initial condition on the boundary (z=0): $A = A_o \cdot \exp(-2\ln(2) t^2/T^2)$. Chirped mirrors are implemented after each nonlinear stage. The mirrors produce a correction of spectral phase and pulse shortening. In the simple case, it corrects only quadratic component of the phase:

$$A_c(t) = F\left(e^{-\frac{i\alpha\omega^2}{2}} F^{-1}(A_{out}(t,L))\right)$$

Here $A_{out}$ and $A_c$ – the amplitudes of the pulse at the output of the nonlinear element and after the recompression, F and $F^{-1}$ are direct and inverse Fourier transform, α - parameter of Group Velocity Dispersion of chirped mirrors.

In order to demonstrate the potential of Thin-Film-Compressor, we use the following initial beam parameters: pulse duration T=27 fs, energy 27 Joules, central wave length 800 nm, flat-top transverse intensity distribution with diameter 160 mm. The thicknesses of the first and second nonlinear elements are 0.5 mm and 0.1 mm. The cubic nonlinearity parameter γ=3.35·$10^{-7}$ $cm^2$/GW, $k_2$=36.7 $fs^2$/mm. The fundamental peak intensity is 4.7 TW/$cm^2$, after the first and second stages with temporal recompression procedure 16.6 TW/$cm^2$ and 43 TW/$cm^2$ at pulse durations 6.4 fs and 2.1 fs correspondingly. The accumulated B integral inside the first and second nonlinear elements are 6.1 and 4.4. The values of B-integral are permissible and small-scale self-focusing can be suppressed in accordance with the technique presented in [9]. The results of numerical simulations : the spectral and temporal intensity profiles are presented on the fig.4.

The proposed technique gives opportunity to compress initially Fourier transform limited pulses and increase peak power by one order of magnitude with help of only passive optical components. More other, the numerical simulations demonstrate the Thin-Film-Compressor doesn't change the transversal intensity distribution significantly. Also, it is necessary to underline the main advantage of the compressor - the possibility to implement it for high energy and super power laser systems.

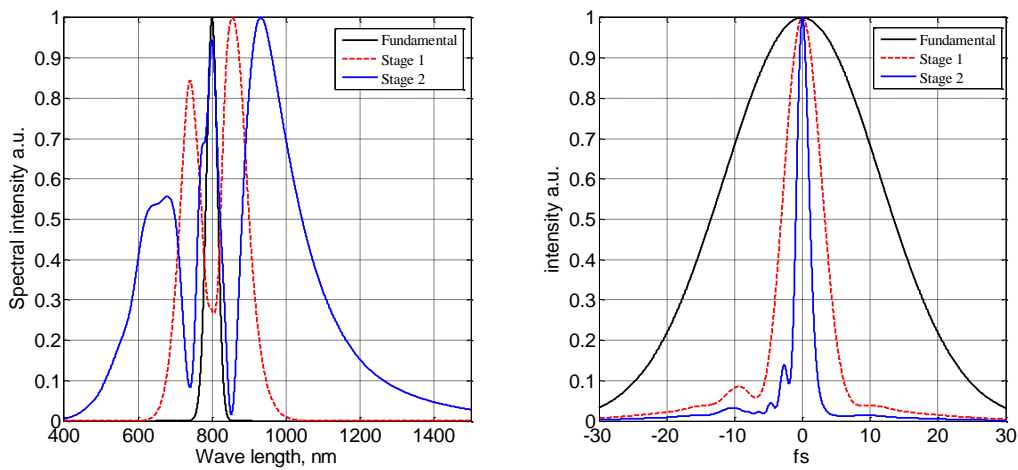

Fig. 4 shows the successive spectra and pulse durations corresponding to the laser out put, after the first stage and second stage. After the first stage the pulse 6.4fs, after the second stage the pulse is shrunk to 2.1fs

## Getting into the Relativistic $\lambda^3$ regime.

This result becomes extremely relevant to the so called Relativistic $\lambda^3$ regime [10] where relativistic few cycle pulses is focused on one $\lambda^2$ – Fig. 5a. The relativistic mirror is not planar and rather deforms due to the indentation created by the focused Gaussian beam. As it moves relativistically in and out and sideways, the reflected beam is broadcasted in specific directions and provide an elegant way to isolate an individual pulse Fig. 5b, c In the relativistic regime Naumova et al. [10] predicts a pulse duration $T$ –compressed by the relativistic mirror - scaling like $T=600$ (*attosecond*)/$a_0$ Fig.6 Here $a_0$ is the normalized vector potential, which is unity at $10^{18}$W/cm$^2$ and scales as the square root of the intensity. Similar results are predicted by the Pukhov' group [11]. For intensity of the order of $10^{22}$W/cm$^2$ the compressed pulse could be of the order of only a few attoseconds even zepto second. Naumova et al [12] have simulated the generation of thin sheets of electrons of few nm thickness, much shorter than the laser period. It opens the prospect for X and gamma coherent scattering with good efficiency. A similar concept called 'relativistic flying mirror' has been demonstrated [13], using a thin sheet of accelerated electrons. Reflection from this relativistic mirror will lead to high efficiency, pulse compression.

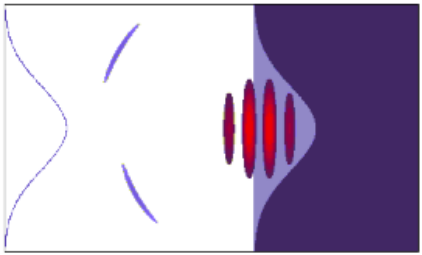

Fig. 5a Interaction of few cycle pulse in the relativistic λregime. It shows the shaped mirror created by the enormous light pressure. In this time scale only the electrons have the time to move. The ions are to slow to follow.

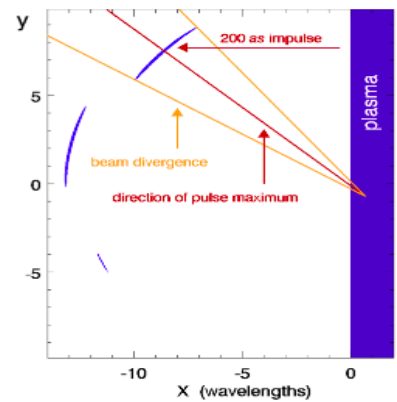

5b The reflection of an ultra relativistic pulse by a high Z target will broadcast the beam in specific way. The pulse is compressed by a factor proportional to a0. The pulses will be easily isolated.

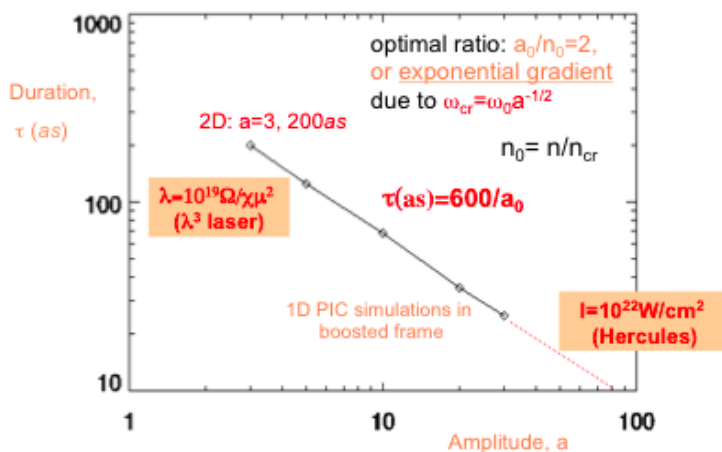

Fig.6 Shows the pulse duration as function of $a_0$, the normalized vector potential.
The expression of the pulse duration is derived to be 600as/a0.
For $a_0$ of the order o 1000, pulse duration of 600zs could be achieved.

**Physics of Vacuum Nonlinearity and Pulse Compression in Vacuum**

As the pulse is compressed into extremely short duration, a modest efficiency could produce sizable nonlinearities in vacuum, although the value of $n_2$ is 18 orders of magnitude smaller than a typical optical transparent medium like glass. The critical power is inversely proportional to the square of the frequency and the vacuum critical power is $10^{24}$W at 1.0 μm [14]. It should be 6 orders of magnitude less for one attosecond pulse, or $10^{18}$W for 1keV X-rays. Under this condition the vacuum critical power could be attained with a single joule. For a 10PW laser with 250J input energy it corresponds to only .4% efficiency. It is quite fascinating to imagine that filaments in vacuum analogous to those produced in air [15] could be generated. Their sizes would be limited by "vacuum breakdown" or pair creation as the intensity approaches $10^{29}$W/cm$^2$ corresponding to a filament of $10^{-5}$ cm diameter.

**Conclusion:** Modern high peak power laser producing PW and 10PW pulses with top hat distribution, combined with Thin Film Compressor will be capable to produce 100PW with single femtosecond duration in the form of ultralativistic $\lambda^3$ pulses. It is predicted that their interaction with solid will generate attosecond or even zeptosecond multi exawatt pulses opening the door to nonlinear optics of vacuum.

**References:**
1. G.A. Mourou and T. Tajima, More intense Shorter Pulse, SCIENCE VOL 331 , 7 , p 41, JANUARY (2011)
2. D. Grischkowsky, A, C. Ballant, Optical pulse compression based on enhanced frequency chirping, Applied Physics Letters, 41, 1982
3. W. N. Knox, R. L. Fork, M. C. Downer, D. A. B. Miller, D. S. Chemla, C. V. Shank, A. C.Gossard, and W. Wiegmann, Phys. Rev. Lett. 54, 1306 (1985).
4. Nisoli, M., De Silvestri,S. & Svelto, O. Generation of high energy 10 fs pulses by a new pulse compression technique. *Appl. Phys. Lett*. 68, 2793-2795 (1996).
5. Nisoli, M., De Silvestri, S., Svelto, O., Szipöcs, R., Ferencz, K., Spielmann, Ch., Sartania, S. & Krausz, F. Compression of High-Energy Laser Pulses below 5fs, *Opt. Letters* 22, 522 (1997).
6. Rolland C and Corkum P.B. JOSA B, Vol. 5, Issue 3, pp. 641-647 (1988)
7. A. K. Potemkin, M. A. Martyanov, M. S. Kochetkova, and E. A. Khazanov, "Compact 300 J/ 300 GW frequency doubled neodimium glass laser. Part I: Limiting power by self-focusing.," IEEE Journal of Quantum Electronics **45**, 336-344 (2009).
8. S. A. Akhmanov, V. A. Vysloukh, and A. S. Chirkin, *Optics of Femtosecond Laser Pulses* (American Institute of Physics; 1992 edition (August 1, 1997), 1992).
9. Sergey Mironov, Vladimir Lozhkarev, Grigory Luchinin, Andrey Shaykin, Efim Khazanov, "Suppression of small-scale self-focusing of high-intensity femtosecond radiation", Appl. Phys. B, 113, pp. 147-151 (2013).
10. Naumova, N. M., Nees, J. A., Sokolov, I. V., Hou, B. & Mourou, G. Relativistic generation of isolated attosecond pulses in a $\lambda^3$ focal volume. *Phys. Rev. Lett*. **92**, 063902 (2004).
11. D. an der Brügge and A. Pukhov, Phys. Plasmas 17, 033110 (2010)
12. Naumova, N., Sokolov, I., Nees, J., Maksimchuk, A., Yanovsky, V., & Mourou, G. Attosecond Electron Bunches, *Phys. Rev. Lett.* **93**, 195003 (2004)
13. Esirkepov, T., Borghesi, M., Bulanov, S. V., Mourou, G., & Tajima, T. Highly efficient relativistic ion generation in the laser-piston regime. *Phys. Rev. Lett.* **92**, 175003 (2004)
14. Mourou, G. A., Tajima, T., & Bulanov, S. V. Optics in the relativistic regime. *Rev. Mod. Phys.* **78**, 309-371 (2006).
15. Braun, A., Korn, G., Liu, X., Du, D., Squier, J., & Mourou, G. Self-Channeling of High-Peak-Power Femtosecond Laser Pulses in Air. *Opt. Lett.* **20**, 73-75 (1995).